# INTERFERENCE MINIMIZATION IN PHYSICAL MODEL OF WIRELESS NETWORKS

## Hakob Aslanyan

*Abstract: Interference minimization problem in wireless sensor and ad-hoc networks is considered. That is to assign a transmission power to each node of a network such that the network is connected and at the same time the maximum of accumulated signal straight on network nodes is minimum. Previous works on interference minimization in wireless networks mainly consider the disk graph model of network. For disk graph model two approximation algorithms with $O(\sqrt{n})$ and $O((opt \ln n)^2)$ upper bounds of maximum interference are known, where $n$ is the number of nodes and $opt$ is the minimal interference of a given network. In current work we consider more general interference model, the physical interference model, where sender nodes' signal straight on a given node is a function of a sender/receiver node pair and sender nodes' transmission power. For this model we give a polynomial time approximation algorithm which finds a connected network with at most $O((opt \ln n)^2 / \beta)$ interference, where $\beta \geq 1$ is the minimum signal straight necessary on receiver node for successfully receiving a message.*

*ACM Classification Keywords: C.2.1 Network Architecture and Design - Network topology, G.2.2 Graph Theory - Network problems.*

## Introduction

We consider interference minimization problem in energy limited wireless networks (wireless sensor and ad-hoc networks) where recharging or changing the energy sources of nodes is not feasible and sometimes due to environmental conditions not possible. In such networks it is important to consider the minimization of energy consumption of algorithms running on network nodes. By decreasing energy consumption we increase nodes operability time and as a result networks' lifetime. In different wireless sensor network (WSN) applications definition of networks' lifetime may be different (till all the nodes are alive, network is connected, given area is monitored by alive nodes, etc). In current work we tend to decrease energy consumption of nodes by decreasing the maximum interference of network algorithmically. Wireless communication of two nodes which is experiencing the third one is called interference. High interference on a receiver node (high value of accumulated signal straights on a node) makes difficulty to determine and accept the signals dedicated to it, this makes necessity for sender node to retransmit the signal until it is successfully accepted by receiver node, which is extra energy consumption and should be avoided.

## Interference Minimization in Disk Graph Model of Wireless Networks

Consider a set of spatially distributed nodes, where each node equipped with radio transmitter/receiver and the power of nodes' transmitter is adjustable between zero and nodes' maximum transmission level. In disk graph model of network assumed that by fixing a transmission power for a node we define a transmission radius/disk of a node, i.e. the transmitted signal is reachable and uniform in any point of transmission disk of node and is zero outside of it. In this model two nodes considered connected if they are covered by each others transmission disks and interference on a given node defined as the number of transmission disks including that node. The overall interference of network is the maximum interference among all the nodes forming the network. The main weakness of disk graph model is the assumption that the radio coverage area is a perfect circle.

*Assigning a transmission powers to a given set of spatially distributed nodes such that nodes form a connected network with assigned transmission powers while the interference of network is minimal called interference minimization problem in wireless networks.*

One particular case of interference minimization problem described above is studied in [Rickenbach, 2005]. Authors considered the problem in one dimensional network, where all the nodes are distributed along the straight line, and named it a highway-model. For this model they showed that intuitive algorithm, which connects each node with its closest left and right nodes, can give a bad performance. An example of network where intuitive algorithm has worst performance is the exponential node chain, where distance between two consecutive nodes grows exponentially ( $2^0, 2^1, \ldots, 2^{n-1}$ ). They also gave two algorithms for one dimensional case of interference minimization problem. The first algorithm, for a given set of distributed nodes, finds a connected network with at most $O(\sqrt{\Delta})$ interference where $\Delta$ is interference of uniform radius network under consideration and is $O(n)$ in some network instances. The second one is an approximation algorithm with $O(\sqrt[4]{\Delta})$ approximation ratio. By applying computational geometry and $\varepsilon$-net theory to ideas given in [Rickenbach, 2005], [Halldorsson, 2006] proposes a algorithm which gives $O(\sqrt{\Delta})$ interference bound for maximum interference in two and $O(\sqrt{\Delta log \Delta})$ for any constant dimensional network. Authors of [Aslanyan, 2010] give iterative algorithm based on linear program relaxation techniques which guaranties $O((opt \ln n)^2)$ interference bound for networks of $n$ nodes, $opt$ here is the optimal interference value for given instance of network. Logarithmic lower bound for interference minimization problem in disk graph model of networks under the general distance function is proven in [Bilo, 2006] by reducing minimum set cover to minimum interference problem.

### Interference Minimization in Physical Model of Wireless Networks

Again, consider a set of spatially distributed wireless nodes, where each node has a radio transmitter/receiver with adjustable power level. In physical model of wireless networks we refuse the assumption that the signal coverage of a node is a perfect circle and assume that the signal straight on any given point (node) of network is a function of sender node, the node in question and the level of transmitted signal. In this model we are also given a constant $\beta$ which is a signal acceptance threshold, i.e. it assumed that receiver node accepts the signal if it's straight is at least $\beta$. By this mean two nodes considered connected if their signals' straights are at least $\beta$ on each other. Interference on a given node defined as a sum of signal straights on that node and interference of networks is the maximum interference among all the nodes forming the network.

The disk graph model can be deduced from physical model if we consider a signal straight function which for every node and its transmission level draws a disk and outputs a positive constant for every node within that disk and zero for the rest. Another example of signal straight function is $f(u,v,\xi) = \xi/d(u,v)^\alpha$ where $u$ and $v$ are sender and receiver nodes respectively, $\xi$ is the transmission power of $u$, $\alpha \in [2,6]$ is the path lost exponent and $d(u,v)$ is the distance between nodes $u$ and $v$ [Pahlavan, 1995].

Interference minimization problem defined in a same way as for disk graph model.

*Assign a transmission powers to a given set of spatially distributed wireless nodes such that nodes form a connected network with assigned transmission powers and the interference of network is minimal.*

Our result is a deterministic polynomial time algorithm for interference minimization problem in wireless networks under the physical model of wireless networks in consideration, which for given network of $n$ wireless nodes finds a connected network with at most $O((opt \ln n)^2/\beta)$ interference.

## Formal Definitions

Consider a set $V$ of $n$ wireless nodes spatially distributed over a given area where nodes have adjustable transmission power and it can be fixed between zero and nodes' maximum transmission power. For any node $u \in V$ denote the range of feasible transmission powers by $R_u = [0, \xi_u^{max}]$, where $\xi_u^{max}$ is the maximum transmission power for node $u$, and define a signal straight function $\phi_u : V \times R_u \to R^+$ where $\phi_u(v, \xi)$ is the signal straight of node $u$ on node $v$ when $u$ uses the transmission power $\xi$. We assume that the signal straight function satisfies to following conditions

1. for any $\xi_1, \xi_2 \in R_u$, from $\xi_1 \geq \xi_2$ it follows that $\phi_u(v, \xi_1) \geq \phi_u(v, \xi_2)$
2. for given $\eta \in R^+$ it is easy to find a $\xi \in R_u$ (if exists) such that $\phi_u(v, \xi) = \eta$

Suppose that for any node $u$ the suitable transmission power $\xi_u$ is fixed, then any two nodes $u$ and $v$ considered connected if $\phi_u(v, \xi_u) \geq \beta$ and $\phi_v(u, \xi_v) \geq \beta$ where $\beta \geq 1$ is the signal acceptance threshold of network. Interference on a given node $u$ is the accumulated signal straight of all the nodes forming the network $I(u) = \sum_{v \in V \setminus \{u\}} \phi_v(u, \xi_v)$ and $I(V) = \max_{v \in V} I(V)$ is the overall network interference. At this point interference minimization problem can be formulated as follows:

*Given a spatially distributed set of wireless nodes, assign a suitable transmission power to each node such that the network is connected and the interference of network is minimal.*

This is the formulation of interference minimization problem by transmission power assignment.

Consider a network graph $G = (V, E)$ where $E = \{(u,v) \mid u,v \in V, \phi_u(v, \xi_u^{max}) \geq \beta, \phi_v(u, \xi_v^{max}) \geq \beta\}$ i.e. in graph $G$ two vertexes/nodes are incident if their maximum transmission powers are enough for communicating with each other. By this mean interference minimization problem is formulated as follows.

*For a given network graph $G = (V, E)$ find a connected spanning subgraph $H = (V, E')$ such that the interference of network computed by the selected set of edges is minimal.*

Formally, having the subgraph $H = (V, E')$ it is correct to further extract transmission power for any node $u$ as a minimum power such that $u$ can communicate with all of its neighbors in $H$, $\xi_u = min_\xi \{\xi \mid \phi_u(v, \xi) \geq \beta \text{ for all } v \text{ that } (u,v) \in E'\}$, which avoids unnecessary interference.

## Set Covering and Interference Minimization

In the classical set cover problem a set $S$ and a collection $C$ of subsets of $S$ are given, it is required to find a minimum size sub collection $C'$ of $C$ such that the union of sets of $C'$ is $S$. In a decision version of set cover problem a positive integer $k$ is given and the question is if it is possible to choose at most $k$ subsets from collection $C$ such that the union of chosen sets is $S$. It is well known that decision version of set cover problem is NP-complete and in polynomial time the optimal solution can not be approximated closer than with a logarithmic factor [Johnson, 1974]. Several variants of set cover problem have been studied [Kuhn, 2005; Garg, 2006; Demaine, 2006; Guo, 2006; Mecke, 2004; Ruf, 2004; Aslanyan, 2003].

Being motivated by interference minimization problem in cellular networks the minimum membership set cover (MMSC) problem has been investigated in [Kuhn, 2005]. In MMSC a set $S$ and a collection $C$ of subsets of $S$ are given, it is required to find a subset $C'$ of $C$ such that the union of sets in $C'$ is $S$ and the maximum covered element of $S$ is covered by as few as possible subsets from $C'$. In a decision version of MMSC problem a positive integer $k$ is given and the question is if it is possible to choose a sub collection of $C$ such that the union of chosen sets is $S$ and each element of $S$ is covered by at most $k$ different subsets. [Kuhn, 2005] Contains the proofs of NP-completeness of decision version of MMSC problem and non-approximability of MMSC optimization problem by factor closer than $O(\ln n)$ unless $NP \subset TIME(n^{O(\log \log n)})$. Also, by using the linear program relaxation and randomized rounding techniques, [Kuhn, 2005] gives a polynomial time algorithm, which approximates the optimal solution of MMSC with logarithmic factor $O(\ln n)$.

Minimum partial membership partial set cover (MPMPSC) problem has been proposed in [Aslanyan, 2010] and used for developing interference minimization algorithm for wireless networks (disk graph model under consideration). In MPMPSC a set $S = S_1 \cup S_2$, consisting of two disjoint sets $S_1$ and $S_2$, along with collection $C$ of subsets of $S$ are given, it is required to find a sub collection $C'$ of $C$ such that the union of sets in $C'$ contains all the elements of $S_1$ and the maximum covered element of $S_2$ is covered by as few as possible subsets from $C'$. In a decision version of MPMPSC problem a positive integer $k$ is given and the question is if it is possible to choose a sub collection of $C$ such that the union of chosen sets contains all the elements of $S_1$ and each element of $S_2$ is covered by at most $k$ different subsets. It is known that the decision version of MPMPSC problem is NP-Complete and that the deterministic polynomial time algorithm exists which approximates the optimal solution of optimization version of MPMPSC by logarithmic factor $O(\log(max\{|S_1|,|S_2|\}))$ which asymptotically matches the lower bound [Aslanyan, 2010]. The approximation algorithm for MPMPSC is achieved by applying the same techniques which has been applied in [Kuhn, 2005] for solving the MMSC.

Being motivated by interference minimization problem in physical model of wireless networks we consider a weighted minimum partial membership partial set cover (WMPMPSC) problem which is a generalization of MPMPSC. In WMPMPSC a set $S = S_1 \cup S_2$, consisting of two disjoint sets $S_1$ and $S_2$, along with collection $C$ of subsets of $S$ are given. In each subset from $C$ the elements of $S_2$ have weights in $[0,1]$. The same element of $S_2$ may have a different weights in different sets of $C$. It is required to find a sub collection $C'$ of $C$ such that the union of sets in $C'$ contains all the elements of $S_1$ and the accumulated, among the subsets of $C'$, weight of a node which has the maximum accumulated weight, is as small as possible. In a decision version of WMPMPSC problem a positive number $k$ is given and the question is if it is possible to choose a sub collection of $C$ such that the union of chosen sets contains all the elements of $S_1$ and the accumulated, among the chosen sets, weight of each node is at most $k$. It is easy to see that in WMPMPSC we get a instance of MPMPSC when each node has a weight 1 in all the sets of $C$. This last statement proves the NP-Completeness of the decision version of WMPMPSC and the logarithmic lower bound for optimization version of the problem.

## LP Formulations

Let $C'$ denote a subset of the collection $C$. To each subset $C_j \in C$ we assign a variable $x_j \in \{0,1\}$ such that $x_j = 1 \Leftrightarrow C_j \in C'$. For $C'$ to be a set cover for $S$, it is required that for each element $u \in S$ at least one set $C_j$ with $u \in C_j$ is in $C'$. Therefore, $C'$ is a set cover for $S$ if and only if for all $u \in S$ it holds that $\sum_{C_j \ni u} x_j \geq 1$. Let $z$ is the maximum membership over all the elements caused by the sets in $C'$. Then for all $u \in S$ it follows that $\sum_{C_j \ni u} x_j \leq z$. Then the integer linear program $IP_{MMSC}$ of MMSC problem can be formulated as:

$$\text{minimize} \quad z$$
$$\text{subject to} \quad \sum_{C_j \ni u} x_j \geq 1, \qquad u \in S \qquad (1)$$
$$\sum_{C_j \ni u} x_j \leq z, \qquad u \in S \qquad (2)$$
$$x_j \in \{0,1\}, \qquad C_j \in C \qquad (3)$$

Integer linear program $IP_{MPMPSC}$ of MPMPSC would be:

$$\text{minimize} \quad z$$
$$\text{subject to} \quad \sum_{C_j \ni u} x_j \geq 1, \qquad u \in S_1 \qquad (4)$$
$$\sum_{C_j \ni u} x_j \leq z, \qquad u \in S_2 \qquad (5)$$
$$x_j \in \{0,1\}, \qquad C_j \in C \qquad (6)$$

After introducing the weight function $w: C \times S_2 \to [0,1]$, where $w(C_j, u)$ is the weight of $u$ in subset $C_j$, the integer linear program $IP_{WMPMPSC}$ of WMPMPSC can be formulated as:

$$\text{minimize} \quad z$$
$$\text{subject to} \quad \sum_{C_j \ni u} x_j \geq 1, \qquad u \in S_1 \qquad (7)$$
$$\sum_{C_j \ni u} x_j w(C_j, u) \leq z, \qquad u \in S_2 \qquad (8)$$
$$x_j \in \{0,1\}, \qquad C_j \in C \qquad (9)$$

By applying randomized rounding technique to $IP_{MMSC}$ with relaxation of constraints $(3)$, [Kuhn, 2005] gives a deterministic polynomial time approximation algorithm with $(1 + O(1/\sqrt{z'}))(\ln(n) + 1)$ approximation ratio for MMSC problem, where $z'$ is the optimal solution for $IP_{MMSC}$ relaxation. Later on [Aslanyan, 2010] states that by applying the same randomized rounding technique to $IP_{MPMPSC}$ with relaxation of constraints $(6)$ gives a deterministic polynomial time approximation algorithm with $(1 + O(1/\sqrt{z'}))(\ln(\max\{|S_1|, |S_2|\}) + 1)$ approximation ratio for MPMPSC problem, where $z'$ is the optimal solution for $IP_{MPMPSC}$ relaxation. In current

work we state that the same randomized rounding technique can be applied to $IP_{WMPMPSC}$ with relaxation of constraints (9) to achieve a deterministic polynomial time approximation algorithm with $(1+O(1/\sqrt{z'}))(\ln(\max\{|S_1|,|S_2|\})+1)$ approximation ratio for WMPMPSC problem, where $z'$ is the optimal solution for $IP_{WMPMPSC}$ relaxation. The proof of the last statement is presented in the Appendix of this work. To sum up, we have the following theorem.

**Theorem 1.** *For WMPMPSC problem, there exists a deterministic polynomial-time approximation algorithm with an approximation ratio of* $O(log(max\{|S_1|,|S_2|\}))$ [1]

## Approximation Algorithm for Interference Minimization in Physical Model of Wireless Networks

Algorithm takes a network graph $G = (V, E)$ with $n$ vertices as an input and after logarithmic number of $k \in O(\log n)$ iterations returns connected subgraph $G_k \subseteq G$ where interference of network corresponding to the graph $G_k$ is bounded by $O((opt \cdot \ln n)^2/\beta)$, where $n = |V|$ is the number of network nodes and $opt$ is the interference of minimum interference connected network.

Algorithm starts the work with the graph $G_0 = (V, E_0)$ where $E_0 = \emptyset$. On the $l^{th}$ iteration, $l \geq 1$, algorithm chooses a subset $F_l \subseteq E \setminus E_{l-1}$ of new edges and adds them to already chosen edge set $E_{l-1} = \cup_{i=1}^{l-1} F_i$. As a consequence of such enlargement of edge set, interference on graph vertices may increase in some value depending on $F_l$. Algorithm finishes the work if the graph $G_l = (V, E_l)$ is connected otherwise goes for the next iteration. Below we present how algorithm chooses the set of edges $F_l \subseteq E \setminus E_{l-1}$ on the $l^{th}$ iteration. Algorithms' quality, i.e the final maximal interference on nodes (its upper estimate) is bounded by the accumulated through the iterations interferences which we try to keep minimal. Let $G_{l-1} = (V, E_{l-1})$ is the graph obtained after the $(l-1)^{th}$ iteration, and has the set of connected components $C(G_{l-1}) = \{C_{l-1}^1, \ldots, C_{l-1}^{k_{l-1}}\}$. Denote by $H_{l-1} \subseteq E \setminus E_{l-1}$ the set of all edges which have their endpoints in different connected components of $G_{l-1}$. On the $l^{th}$ stage of algorithm a subset of $H_{l-1}$ is selected to further reduce the number of connected components which finally brings us to a connected subgraph. In this way we build the collection $T(C(G_{l-1}), H_{l-1})$ of special sets as follows. Starting with $H_{l-1}$ we add to the set $T(C(G_{l-1}), H_{l-1})$ of $l^{th}$ stage specific weighted subsets $T^l(u,v) = \{C_{l-1}^u, C_{l-1}^v\} \cup V$ defined by all $(u,v) \in H_{l-1}$, where $u$ belongs to connected component $C_{l-1}^u$ and $v$ belongs to $C_{l-1}^v$. By selection of $u$ and $v$ we have that $C_{l-1}^u$ and $C_{l-1}^v$ are different. By definition of connectivity nodes $u$ and $v$ can communicate with each other if their signal transmission powers $\xi_{uv}$ and $\xi_{vu}$ satisfy to $\phi_u(v, \xi_{uv}) \geq \beta$ and $\phi_v(u, \xi_{vu}) \geq \beta$, where $\beta$ is the signal acceptance threshold. To avoid unnecessary energy consumption and to reduce interference it would be right to adjust transmission powers $\xi_{uv}$ and $\xi_{vu}$ such that $\phi_u(v, \xi_{uv}) = \beta$ and $\phi_v(u, \xi_{vu}) = \beta$, this is possible to do because of the second property of the signal straight function $\phi$. Then the noise of the link $(u,v)$ on any node $t$ can be calculated as $w((u,v),t) = \phi_u(t, \xi_{uv}) + \phi_v(t, \xi_{vu})$ which would be the weight $w(T^l(u,v),t)$ of node $t$ in the subset $T^l(u,v)$. And so $T^l(u,v)$ is a composite set which includes two labels for components $C_{l-1}^u$ and $C_{l-1}^v$ and all the vertices in $V$ along with the weights, which are the interference increase on nodes if the edge $(u,v)$ is selected as a communication link. In terms of WMPMPSC the labels of connected components will compose the set $S_1$ and weighted $V$ will be the set $S_2$.

---

[1]See the Appendix $A$ for the proof.

Figure 1 demonstrates connected components that are input to the stage $l$, and the set $H_{l-1}$ of all cross component edges.

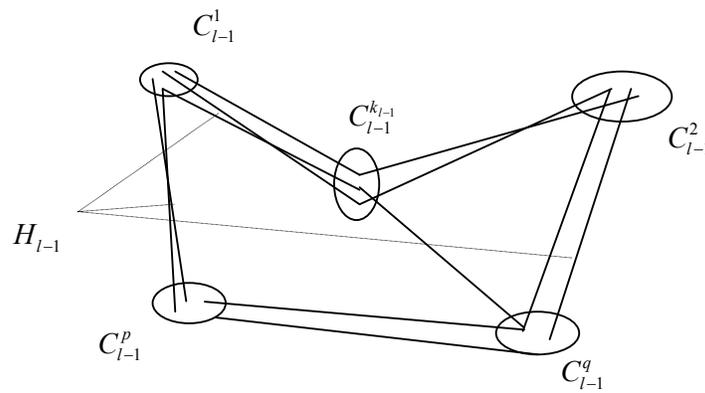

Figure 1: Connected components that are input to the $l$-th stage of the algorithm

After constructing $T(C(G_{l-1}), H_{l-1})$ we normalize the weights of elements by dividing all the weights by the maximum weight $w_{max} = \max_{t,(u,v) \in H_{l-1}} w((u,v),t)$ and solve the WMPMPSC on the set $C(G_{l-1}) \cup V$ and collection of subsets $T(C(G_{l-1}), H_{l-1})$, where condition for elements of $C(G_{l-1})$ is *to be covered* and for elements of $V$ is *to have minimum accumulated weight*. Finally, based on the solution $W(C(G_{l-1}), H_{l-1}) \subseteq T(C(G_{l-1}), H_{l-1})$ of WMPMPSC we build the set $F_l$ of network graph edges, selected at the $l^{th}$ iteration of algorithm by adding to $F_l$ all the edges $(u,v) \in H_{l-1}$ such that $T^l(u,v) \in W(C(G_{l-1}), H_{l-1})$ and multiply all the weights by $w_{max}$ to receive the real interference increase.

### Algorithm performance

**Theorem 2.** *On each iteration of algorithm the number of connected components is being reduced at least by factor of two, which bounds the total number of iterations by $O(\log n)$.*

*Proof.* For each connected component $C_{l-1}^u \in C(G_{l-1})$ of graph $G_{l-1}$ the solution $W(C(G_{l-1}), H_{l-1})$ of WMPMPSC solved at $l^{th}$ iteration contains at least one set $T^l(u,v) \in W(C(G_{l-1}), H_{l-1})$ such that $C_{l-1}^u \in T^l(u,v)$ (as $W(C(G_{l-1}), H_{l-1})$ is a cover for the set $C(G_{l-1})$). And as each set $T^l(u,v) \in W(C(G_{l-1}), H_{l-1})$ contains exactly two connected components, then by adding the edge $(u,v)$ to our solution, we merge those two connected components into one (connecting by the edge $(u,v)$). So every connected component merges with at least one other component, which reduces the number of connected components at least by factor of $2$.

**Lemma 1.** *Network corresponding to the graph $G^l = (V, F_l)$, where $F_l$ is the edge set obtained on the $l^{th}$ iteration of algorithm, has interference in $O((opt^2 \cdot \ln n)/\beta)$.*

*Proof.* Consider the set of connected components $C(G_{l-1}) = \{C_{l-1}^1, \ldots, C_{l-1}^{k_{l-1}}\}$ of $l^{th}$ iterative step of algorithm. Let $E_{opt}$ is the set of the edges of some interference optimal connected network for our problem (edges of connected network with optimal interference $opt$). Then there is a subset $E_{opt}^l \subseteq E_{opt}$ which spans connected components $C(G_{l-1})$ and the network of the graph $G_{opt}^l = (V, E_{opt}^l)$ has interference not exceeding the $opt$.

**Fact 1.** *The maximal vertex interference due to the spanner $E_{opt}^l$ of $C(G_{l-1})$ is at most $opt$.*

Now let us build the set collection $T_{opt}(C(G_{l-1}), E_{opt}^l) = \{T^l(u,v)/(u,v) \in E_{opt}^l\}$.

**Fact 2.** *$T_{opt}(C(G_{l-1}), E_{opt}^l)$ is a sub collection of $T(C(G_{l-1}), H_{l-1})$ built on the $l^{th}$ iteration of algorithm and is a cover for $C(G_{l-1})$, i.e. $T_{opt}(C(G_{l-1}), E_{opt}^l)$ is a solution for the WMPMPSC problem, with some value $z^*$, solved on the $l^{th}$ iteration of algorithm, not necessary optimal.* Now consider the matrix $P_{opt}^w$ representing the transmission signals on some node $w$ caused by communication links of $E_{opt}^l$.

$$P_{opt}^w = \begin{pmatrix} P_{u_1 u_1}^w & P_{u_1 u_2}^w & \cdots & P_{u_1 u_j}^w & \cdots & P_{u_1 u_n}^w \\ P_{u_2 u_1}^w & P_{u_2 u_2}^w & \cdots & P_{u_2 u_j}^w & \cdots & P_{u_2 u_n}^w \\ \vdots & \vdots & \ddots & \vdots & \ddots & \vdots \\ P_{u_i u_1}^w & P_{u_i u_2}^w & \cdots & P_{u_i u_j}^w & \cdots & P_{u_i u_n}^w \\ \vdots & \vdots & \ddots & \vdots & \ddots & \vdots \\ P_{u_n u_1}^w & P_{u_n u_2}^w & \cdots & P_{u_n u_j}^w & \cdots & P_{u_n u_n}^w \end{pmatrix}$$

where

$$P_{u_i u_j}^w = \begin{cases} 0, & \text{if } i = j \text{ or } (u_i, u_j) \notin E_{opt}^l \\ \phi_{u_i}(w, \xi_{u_i u_j}), & \text{otherwise} \end{cases}$$

is the signal straight of node $u_i$ on node $w$ when $u_i$ uses the transmission power $\xi_{u_i u_j}$ (communicates with node $u_j$).

**Fact 3.** *and the sum of the matrix elements will give the interference increase we count (the real interference increase is the sum of the maximal elements from each row) on node $w$ by edge set $E_{opt}^l$. Due to the Fact 1 and signal acceptance threshold $\beta$ for any vertex $u_i$ the number of sets $T^l(u_i, v) \in T_{opt}(C(G_{l-1}), E_{opt}^l, w)$ will not* *exceed the* $\lfloor opt/\beta \rfloor$, *in other words the number of non zero elements on each row of matrix* $P_{opt}^w$ *is bounded by* $\lfloor opt/\beta \rfloor$.

**Fact 4.** *The interference increase on node* $w$ *by the edge set* $E_{opt}^l$ *can be calculated as* $\sum_{i=1}^{n} \max_j P_{u_i u_j}^w$ *and due to the Fact 1 it doesn't exceed the* $opt$.

From facts 3 and 4 it follows that the sum of the matrix elements is bounded by $opt^2/\beta$, which means that the optimal value of WMPMPSC problem solved on the $l^{th}$ iteration of algorithm is bounded by $opt^2/\beta$ and therefor by Theorem 1 the interference increase by the edge set $F_l$ is bounded by $O(opt^2 \cdot \ln n/\beta)$.

**Theorem 3.** *The network built by WMPMPSC relaxation algorithm has at most* $O((opt^2 \cdot \ln^2 n)/\beta)$ *interference.*

*Proof.* The proof is in combination of Theorem 2 and Lemma 1.

## Conclusion and Future Work

In current work we considered the interference minimization problem in physical model of wireless networks and proposed a polynomial time approximation algorithm which for a given set of wireless nodes creates a connected network with at most $O((opt \cdot \ln n)^2/\beta)$ interference. In some WSN applications network considered as functional while it is connected, therefore in future works on interference minimization the $k$-connectivity of network should be considered. Also considering the problem in Euclidean spaces, which is a realistic case for WSNs, may give a better approximation ratio.

## Appendix A

Here we show how randomized rounding technique used in [Kuhn, 2005] for solving the $IP_{MMSC}$ can be used for solving $IP_{WMPMPSC}$. This section mostly presents the work of [Kuhn, 2005].

Consider a instance $(S = S_1 \cup S_2, C, w)$ of $IP_{WMPMPSC}$ and the solution vector $\underline{x}'$ and $\underline{z}'$ of $LP_{WMPMPSC}$ relaxation of $IP_{WMPMPSC}$. Consider the following randomized rounding scheme, where an integer solution $\overline{x} \in {0,1}^m$ is computed by setting

$$x_i = \begin{cases} 1, & \text{with probability } p_i := \min\{1, \alpha x'_i\} \\ 0, & \text{otherwise} \end{cases}$$

independently for each $i \in \{1,...,n\}$. Let $A_i$ be the "bad" event that the $i^{th}$ element is not covered.

**Lemma A1.** *The probability that the $i^{th}$ element remains uncovered is*

$$P(A_i) = \prod_{C_j \ni u_i} (1 - p_j) < e^{-\alpha}$$

*Proof.* The proof is in Lemma 1 of [Kuhn, 2005].

Let $B_i$ be the "bad" event that the weight of the $i^{th}$ element is more than $\alpha\beta z'$ for some $\beta \geq 1$.

**Lemma A2.** *The probability that the weight of the $i^{th}$ element is more than $\alpha\beta z'$ is*

$$P(B_i) < \frac{1}{\beta^{\alpha\beta z'}} \cdot \prod_{C_j \ni u_i} (1 + (\beta^{w(j,i)} - 1)p_j) \leq \left(\frac{e^{\beta-1}}{\beta^\beta}\right)^{\alpha z'}$$

*Proof.* We use a Chernoff-type argument. For $t = \ln \beta > 0$, we have

$$P(B_i) = P\left(\sum_{C_j \ni u_i} x_j w(j,i) > \alpha\beta z'\right) = P\left(e^{t \cdot \sum_{C_j \ni u_i} x_j w(j,i)} > e^{t \cdot \alpha\beta z'}\right)$$

$$< \frac{E\left[e^{t \cdot \sum_{C_j \ni u_i} x_j w(j,i)}\right]}{e^{t \cdot \alpha \beta z'}} = \frac{1}{e^{t \cdot \alpha \beta z'}} \cdot \prod_{C_j \ni u_i} \left[p_j e^{t \cdot w(j,i)} + 1 - p_j\right]$$

$$= \frac{1}{\beta^{\alpha \beta z'}} \cdot \prod_{C_j \ni s_i} \left[1 + (\beta^{w(j,i)} - 1) p_j\right] \leq \frac{1}{\beta^{\alpha \beta z'}} \cdot \prod_{C_j \ni s_i} e^{p_j (\beta^{w(j,i)} - 1)}$$

$$\leq \frac{1}{\beta^{\alpha \beta z'}} \cdot \prod_{C_j \ni s_i} e^{p_j (\beta - 1) w(j,i)} = \frac{1}{\beta^{\alpha \beta z'}} \cdot e^{(\beta - 1) \cdot \sum_{C_j \ni u_i} p_j w(j,i)} \leq \left(\frac{e^{(\beta-1)}}{\beta^\beta}\right)^{\alpha z'}$$

The inequality and equality in the second line results by application of the Markov inequality and because of the independence of the $x_j$. The equality and inequality in the third line hold because $t = \ln \beta$ and $1 + x \leq e^x$. For the inequalities in the last line we apply $\beta^x - 1 \leq (\beta - 1)x$ for $\beta \geq 1$, $x \in [0,1]$ and $\sum_{C_j \ni u_i} p_j w(j,i) \leq \alpha z'$.

Denote the probability upper bounds given by Lemmas $A1$ and $A2$ by $\overline{A}_i$ and $\overline{B}_i$:

$$\overline{A}_i := \prod_{C_j \ni s_i} (1 - p_j) \quad \text{and} \quad \overline{B}_i := \frac{1}{\beta^{\alpha \beta z'}} \cdot \prod_{C_j \ni u_i} (1 + (\beta^{w(j,i)} - 1) p_j).$$

In order to bound the probability for any "bad" event to occur, we define a function $P$ as follows

$$P(p_1, \ldots, p_m) := 2 - \prod_{i=1}^{n} (1 - \overline{A}_i) - \prod_{i=1}^{n} (1 - \overline{B}_i).$$

**Lemma A3.** *The probability that any element is not covered or has a weight more than $\alpha \beta z'$ is upper-bounded by $P(p_1, \ldots, p_m)$:*

$$P\left(\bigcup_{i=1}^{n} A_i \cup \bigcup_{i=1}^{n} B_i\right) < P(p_1, \ldots, p_m).$$

*Proof.* The proof is in Lemma 3 of [Kuhn, 2005].

The following shows that if $\alpha$ and $\beta$ are chosen appropriately, $P(p_1, \ldots, p_m)$ is always less than $1$.

**Lemma A4.** *When setting $\alpha = \ln(\max\{|S_1|, |S_2|\}) + 1$, then for $\beta = 1 + \max\{\sqrt{3/z'}, 3/z'\}$, we have $P(p_1, \ldots, p_m) < 4/5$.*

*Proof.* The proof is in Lemma 4 of [Kuhn, 2005].

Lemmas A1 – A4 lead to the following randomized algorithm for the WMPMPSC problem. As a first step, the linear program $LP_{WMPMPSC}$ has to be solved. Then, all $x'_i$ are rounded to integer values $x_i \in \{0,1\}$ using the described randomized rounding scheme with $\alpha = \ln(\max\{|S_1|, |S_2|\}) + 1$. The rounding is repeated until the solution is feasible (all elements are covered) and the weight of the integer solution deviates from the fractional weight $z'$ by at most a factor $\alpha\beta$ for $\beta = 1 + \max\{\sqrt{3/z'}, 3/z'\}$. Each time, the probability to be successful is at least $1/5$ and therefore, the probability of not being successful decreases exponentially in the number of trials.

We will now show that $P(p_1, \ldots, p_m)$ is a pessimistic estimator and that therefore, the algorithm described above can be derandomized. That is, $P$ is an upper bound on the probability of obtaining a "bad" solution, $P < 1$ ($P$ is a probabilistic proof that a "good" solution exists), and the $p_i$ can be set to $0$ or $1$ without increasing $P$. The first two properties follow by Lemmas A3 and A4, the third property is shown by the following lemma.

Lemma A5.

For all $i$, either setting $p_i$ to $0$ or setting $p_i$ to $1$ does not increase $P$:

$$P(p_1, \ldots, p_m) \geq \min\{P(\ldots, p_{i-1}, 0, p_{i+1}, \ldots), P(\ldots, p_{i-1}, 1, p_{i+1}, \ldots)\}$$

*Proof.* The proof is in Lemma 5 of [Kuhn, 2005].

Lemmas A3, A4 and A5 lead to an efficient deterministic approximation algorithm for the WMPMPSC problem. First, the linear program $LP_{WMPMPSC}$ has to be solved. The probabilities $p_i$ are determined as described above. For $\alpha$ and $\beta$ as in Lemma A4, $P(p_1, \ldots, p_m) < 4/5$. The probabilities $p_i$ are now set to $0$ or $1$ such that $P(p_1, \ldots, p_m)$ remains smaller than $4/5$. This is possible by Lemma A5. When all $p_i \in \{0,1\}$, we have an integer solution for $IP_{WMPMPSC}$. The probability that not all elements are covered or that the weight is larger than $\alpha\beta z'$ is smaller than $P < 4/5$. Because all $p_i$ are $0$ or $1$, this probability must be $0$. Hence, the computed $IP_{WMPMPSC}$-solution is an $\alpha\beta$-approximation for WMPMPSC.

Authors' Information

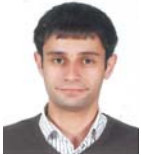

*Hakob Aslanyan – PhD student and research assistant; Computer Science Department, University of Geneva, Battelle Batiment A, route de Drize 7, 1227 Geneva, Switzerland; e-mail: hakob.aslanyan@unige.ch*

*Major Fields of Scientific Research: combinatorial optimization, graph theory, approximation and exact algorithms, hardness of approximation, network design and connectivity*